\documentstyle[epsf,prl,aps,floats]{revtex}

\newcommand{\REV}{}
\newcommand{\VER}{}
\newcommand{\SHMZin}{}
\newcommand{\SHMZout}{}

\newcommand{\elec}{{\it e\/}}
\newcommand{\hole}{{\it h\/}}

\newcommand{\ee}{{\it e-e\/}}
\newcommand{\eh}{{\it e-h\/}}
\newcommand{\hh}{{\it h-h\/}}

\newcommand{\supe}{^{e}}
\newcommand{\suph}{^{h}}

\newcommand{\supeh}{^{eh}}

\newcommand{\supee}{^{ee}}
\newcommand{\suphh}{^{hh}}

\newcommand{\mitDelta}{{\mit\Delta}}
\newcommand{\mitdelta}{{\mit\delta}}
\newcommand{\Deltaeh}{\mitDelta\supeh}

\newcommand{\Deltaee}{\mitDelta\supee}
\newcommand{\Deltahh}{\mitDelta\suphh}

\newcommand{\deltaeh}{\mitdelta\supeh}

\newcommand{\deltaee}{\delta\supee}

\newcommand{\Ne}{N^{e}}
\newcommand{\Nh}{N^{h}}
\newcommand{\Nx}{N^{x}}
\newcommand{\Nd}{N^{d}}

\newcommand{\Ueh}{U\supeh}
\newcommand{\Uee}{U\supee}
\newcommand{\Uhh}{U\suphh}
\newcommand{\Veh}{V\supeh}
\newcommand{\Vee}{V\supee}

\newcommand{\mue}{\mu^{e}}
\newcommand{\muh}{\mu^{h}}

\newcommand{\Vk}{{\bf k}}
\newcommand{\sVk}{{\bf k}}

\newcommand{\nF}{{n_{\rm F}}}

\newcommand{\LPippard}{{L\supeh}}

\newcommand{\Nxopt}{\Nx_{\rm opt}}
\newcommand{\subtot}{_{\rm tot}}
\newcommand{\subopt}{_{\rm opt}}
\newcommand{\subBCS}{_{\rm BCS}}

\begin{document}

\twocolumn[
    \hsize\textwidth\columnwidth\hsize\csname 
    @twocolumnfalse\endcsname

\title{
Strong Enhancement of Superconducting Correlation \\ 
in a Two-Component Fermion Gas
}
\author{
  Masaki {\sc Shigemori}$^{1,}$%
  \cite{address_Shigemori},
  Akira {\sc Shimizu}$^{1,2}$%
  ,
  Tobias {\sc Brandes}$^{1,2,}$%
  \cite{address_Brandes},
  and Jun-ichi {\sc Inoue}$^{1,2}$%
}
\address{
  $^1$Department of Basic Science, University of Tokyo, 
  Komaba, Meguro-ku, Tokyo 153-8902
  \\
  $^2$Core Research for Evolutional
  Science and Technology, JST
}
  
\date{\today}

\maketitle
\begin{abstract}
     We study high-density electron-hole ({\eh}) systems 
\SHMZin
     with the electron density slightly higher than 
\SHMZout
     the hole density.
     We find a new superconducting phase, 
     in which the excess electrons form Cooper pairs 
     moving in an {\eh} BCS phase.
     The coexistence of the {\eh} and {\ee} orders 
     is possible because {\elec} and {\hole} have opposite charges,
     whereas
     analogous phases are impossible 
     in the case of two fermion species
     that have the same charge or are neutral.
     Most strikingly, 
     the {\eh} order enhances the superconducting {\ee} order parameter 
     by more than one order of magnitude
     as compared with that given by the BCS formula,
     for the same value of the effective {\ee} attractive potential
     $\lambda^{ee}$.
     This new phase should be observable in an
     {\eh} system created 
     by photoexcitation in doped semiconductors
     at low temperatures.
\end{abstract}

\vspace*{3ex}
]


It is expected that 
electron-hole ({\eh}) systems created through 
photoexcitation of semiconductors 
exhibit various phases
depending on the material parameters
and 
the densities $\Ne$ and $\Nh$ of electrons 
and holes, respectively~\cite{Keldysh-in-BEC}.
Some of the interesting phases have been successfully 
observed, including the {\eh} liquid~\cite{EHL}
and the Bose--Einstein condensation of 
excitons~\cite{condensation-1,condensation-2}.
These successes
are largely due to
the careful control of both the material parameters
(by choosing a semiconductor) 
and the {\eh} density (through the excitation intensity).
By further increasing the {\eh} density, 
one should observe 
an ``{\eh} BCS phase''
at low temperatures,
which is characterized by a nonzero {\eh} 
order parameter \cite{excitonic_insulator}.
%
Although 
these phases are realized when $\Ne=\Nh$,
one can also use {\em doped semiconductors}.
In this case, one can control
$\Ne$ and $\Nh$ {\em independently\/};
$\Ne - \Nh \equiv \Nx$
through the donor (or acceptor) density $\Nd$, 
and $\Ne + \Nh$ through the excitation intensity \cite{excitonic_insulator}.
The additional parameter $\Nx$ may lead to 
a new quantum phase(s) at low temperatures. 
When $0 < |\Nx| \ll \Ne \approx \Nh$, in particular, 
we may expect a new superconducting phase, 
which we call a 
\REV multiply ordered superconducting (MS) \VER 
phase,
in which doped electrons form Cooper pairs 
moving in the {\eh} BCS state.
On the other hand, 
most of the previous studies on 
superconductivity in the doped 
\SHMZin
{\eh} BCS state (or related systems 
such as doped excitonic insulators and
doped exciton systems) treated
either 
the case where 
the doped electrons are located in a third band 
that is different from the bands of {\eh} pairs
\cite{exciton_mechanism-1,exciton_mechanism-2}, 
or the case $\Nh \ll \Ne$ 
(or equivalently, $\Ne \ll \Nh$)~\cite{previousEI-1,previousEI-2}.
\SHMZout
In these theories, 
the {\eh} BCS condensates (or excitons) work as
polarizable media, which induce a
large attractive interaction between electrons, 
whereas 
the superconducting order parameter $\Deltaee$
as well as the superconducting transition temperature $T_{\rm c}^{ee}$ is
essentially given by substituting $\lambda^{ee}$ 
(effective dimensionless $e$-$e$ attraction) and a cutoff parameter
into the BCS formula or the McMillan formula~\cite{McMillan}.
However, 
we expect the new 
\REV MS \VER
 phase when 
excess electrons (or holes) are doped in the $e$ ($h$) band 
{\em where electrons (holes) forming the {\eh} BCS 
state are located\/},
and when {\em the {\eh} pairing is dominant\/}, i.e.,
when $0 < |\Nx| \ll \Ne \approx \Nh$ and  
$|\Deltaee|, |\Deltahh| \ll |\Deltaeh|$.
Here,
$\Deltaee$ and $\Deltahh$ are the superconducting $e$-$e$
and $h$-$h$ order parameters, respectively, and 
$\Deltaeh$ denotes the $e$-$h$ order parameter.

In this Letter, we explore the possibility of such a new 
\REV MS \VER phase, by studying 
the phase diagram of high-density 
$e$-$h$ systems as a function of $\Nx$.
It is shown that the 
\REV MS \VER
phase,
which has the order parameters 
$0< (|\Deltaee|, |\Deltahh|) \ll |\Deltaeh|$,
can be realized when 
$0 < |\Nx| \ll \Ne \approx \Nh$.
Most strikingly, 
$|\Deltaee|$ is enhanced in the 
\REV MS \VER 
phase by more than one order of magnitude 
in comparison with the value given by 
the BCS or McMillan formula,  
for the same value of $\lambda^{ee}$.

%
%


We assume a three-dimensional, high-density, 
isotropic $e$-$h$ gas at zero temperature, 
so that 
BCS-like mean field approximations
work well~\cite{excitonic_insulator}.
We decompose the interaction Hamiltonian
$H_{\rm int}$ into the short- and 
long-range parts, $H_{\rm int}^{\rm SR}$ 
and $H_{\rm int}^{\rm LR}$, respectively. 
The latter is related to long-range charge fluctuations,
and will be discussed later.
For the moment, we consider a charge-neutral region
of unit volume, 
in which $H_{\rm int}^{\rm LR}$ is irrelevant.
In $H_{\rm int}^{\rm SR}$, 
the {\ee}, {\hh} and {\eh} interaction matrix elements
are renormalized to effective
values, $\Uee$, $\Uhh$ and $\Ueh$, respectively,  
due to 
the screening effect and negative (attractive) contributions
from various bosonic excitations, such as lattice phonons and excitons.
Since the bare value of $\Ueh$ is negative, 
the bosonic excitations make it more negative, 
whereas (basically positive) $\Uee$ and $\Uhh$ are reduced.
Hence, $|\Ueh|$ tends to be larger than $|\Uee|, |\Uhh|$.
It was suggested that $\Uee$ (and $\Uhh$) can be negative 
for some parameter values
\cite{exciton_mechanism-1,exciton_mechanism-2,
previousEI-1,previousEI-2}.
In the {\eh} BCS phase, 
there is an additional negative contribution from 
the Goldstone mode of 
the {\eh} order parameter.
Since there is no reliable method of 
estimating $\Uee$ and $\Uhh$\cite{migdal}, 
%
we here {\em treat them as given parameters}, 
assuming that $\Uee$, $\Uhh<0$ and $|\Ueh| > |\Uee|, |\Uhh|$, 
and explore 
\SHMZin
the phase diagram
as a function of them and 
$\Nx$ ($\geq 0$).
We find that 
the minimal form of $H_{\rm int}^{\rm SR}$ for the 
\REV MS \VER
state 
is 
\begin{eqnarray}
H_{\rm int}^{\rm SR}
 &=& \sum_{\sVk \neq \sVk'} 
   \Ueh_{\sVk\sVk'}
   \big(
    e_{ \sVk \uparrow}^{\dagger} h_{-\sVk\downarrow}^{\dagger}
    h_{-\sVk'\downarrow} e_{\sVk'\uparrow}
    + e_{ \sVk \downarrow}^{\dagger} h_{-\sVk \uparrow}^{\dagger}
      h_{-\sVk'\uparrow} e_{\sVk'\downarrow}
    \big)
    \nonumber \\
 &&+ \sum_{\sVk \neq \sVk'}
   \Uee_{\sVk\sVk'}
    e_{ \sVk \uparrow}^{\dagger} e_{-\sVk\downarrow}^{\dagger}
    e_{-\sVk'\downarrow} e_{\sVk'\uparrow},
  \label{eq:Hsr}
\end{eqnarray}
where 
$e_{\sVk\sigma}$ ($h_{\sVk\sigma}$) denotes the annihilation operator
of $e$ ($h$) with momentum $\Vk$
and spin $\sigma$,
and
\begin{eqnarray}
 \cases{
  \Ueh_{\sVk \sVk'}=-\Veh,\ \Uee_{\sVk \sVk'}=-\Vee
  &
  if $|\xi_{\sVk}|,\, |\xi_{\sVk'}| < \omega_{\rm c}$,
  \cr
  \Ueh_{\sVk \sVk'}=\Uee_{\sVk \sVk'}=0 & otherwise.
 }
 \label{eq:choice_of_Vkk'}
\end{eqnarray}
Here, $\Veh, \, \Vee>0$ are constants, and $\omega_{\rm c}$ ($\ll \mu$) is
a cutoff of the interactions~\cite{different_cutoffs}.
We assume that $\Veh > \Vee$ so that 
$|\Deltaeh|\gg|\Deltaee|$.
Although we do not include an {\hh} interaction $\Uhh$ in ${\cal H}$,
we have confirmed that $\Uhh$ can 
only modify the magnitude of
the pair correlations by a factor of order unity, as long as
$|\Uhh|\lesssim |\Uee|$.
We do not include terms of the form
$e e h h + \mbox{h.c.}$
(such terms 
are important in the case of 
two-band superconductors~\cite{two-band-sc}),
because in $e$-$h$ systems intermediate processes
involving such terms cost a large amount of energy of the order of 
the energy gap $E_{\rm g}$ ($\gg |U^{eh}|$~\cite{excitonic_insulator}).
The total Hamiltonian without $H_{\rm int}^{\rm LR}$ is denoted by $H$, 
and we put   
\begin{eqnarray}
  {\cal H}
  &\equiv&
  H - \mue \hat{N}^{e} - \muh \hat{N}^{h}
  \nonumber\\
 &=&
  \sum_{\sVk\sigma}
  \big[
  (\xi_{\sVk}\!-\nu) e_{\sVk\sigma}^{\dagger} e_{\sVk\sigma}
  \!+\! (\xi_{\sVk}\!+\nu) h_{\sVk\sigma}^{\dagger} h_{\sVk\sigma}
  \big]
  \!+\! H_{\rm SR}.
  \label{eq:start_H}
\end{eqnarray}
Here, 
$\mue \equiv E_{\rm g}/2 + \mu + \nu$ and 
$\muh \equiv E_{\rm g}/2 + \mu - \nu$ ($\nu \geq 0$) are the
chemical potentials of {\elec} and {\hole}, respectively,
which are assumed to have the same energy dispersion
$\Vk^2 / (2m) + E_{\rm g}/2$.
Moreover,
$\hat{N}\supe \equiv \sum_{\sVk\sigma}
e_{\sVk\sigma}^{\dagger} e_{\sVk\sigma}$,
$\hat{N}\suph \equiv \sum_{\sVk\sigma}
h_{\sVk\sigma}^{\dagger} h_{\sVk\sigma}$,
$\xi_{\sVk} \equiv \Vk^2 / (2m) -\mu$,
and we take $\hbar = 1$.

We apply the mean field approximation
that assumes the {\eh} correlation
$
\deltaeh_{\sVk'}
\equiv
\langle h_{-\sVk'\downarrow} e_{\sVk'\uparrow} \rangle
$ 
and the {\ee} correlation
$
\deltaee_{\sVk'}
\equiv
\langle e_{-\sVk' \downarrow} e_{\sVk'\uparrow} \rangle
$.
\SHMZout
We assume that
$\langle h_{-\sVk\downarrow} e_{\sVk\uparrow} \rangle
=\langle h_{-\sVk\uparrow} e_{\sVk\downarrow} \rangle$.
Using eq.\ (\ref{eq:choice_of_Vkk'}),
we find that the order parameters, defined by 
	$
	\Deltaeh_{\sVk}
	   \equiv
	   \sum_{\sVk'}
	   \Ueh_{\sVk \sVk'} \, \deltaeh_{\sVk'}
	$
	and
	$
	  \Deltaee_{\sVk}
	   \equiv
	   \sum_{\sVk'}
	   \Uee_{\sVk \sVk'} \, \deltaee_{\sVk'}
	$,
take simple forms:
$\Deltaeh_{\sVk}=\Deltaeh$ and $\Deltaee_{\sVk}=\Deltaee$
if $|\xi_{\sVk}|<\omega_{\rm c}$;
$\Deltaeh_{\sVk}=\Deltaee_{\sVk}=0$ otherwise.
Here, $\Deltaeh$ and $\Deltaee$ are constants, which are taken
to be real without loss of generality.
We then diagonalize ${\cal H}$, 
and obtain the self-consistent equations~\cite{Moulopoulos-Ashcroft}.
The system is characterized by
$\Nx$ and dimensionless effective coupling constants 
$\lambda\supeh \equiv \nF \Veh$, 
$\lambda\supee \equiv \nF \Vee$,
where $\nF$ is the density of states per spin at the Fermi surface.

\begin{figure}[t]
 \vspace*{0.2cm}\hspace*{0.4cm}
 \epsfxsize=7.5cm
 \epsfbox{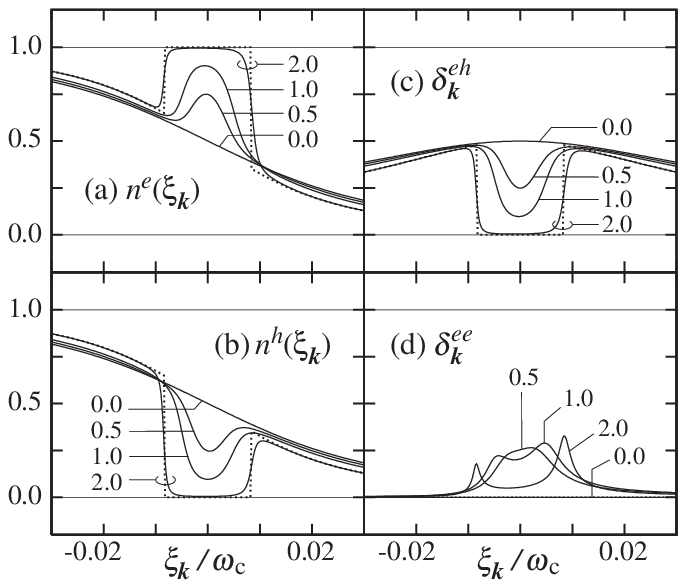}
 \vspace*{0.2cm}
 \caption{
 (a) $n^e(\xi_{\sVk})$,
 (b) $n^h(\xi_{\sVk})$,
 (c) $\deltaeh_{\sVk}$, 
 and 
 (d) $\deltaee_{\sVk}$
 plotted against $\xi_{\sVk}$
 for solutions III (PU state; denoted by dotted lines) 
 and V 
 (\REV MS \VER 
 state; solid lines),
 for $\Nx$ ranging from 0.0 to 2.0 
 in units of $\Nxopt$.
 We take $\lambda\supeh=0.25$ and $\lambda\supee=0.10$.
 }
 \label{fig:pair_corr_vs_xi}
\end{figure}
We have solved the self-consistent equations numerically
and found five solutions, which we denote by I--V:

\noindent
I: $|\Deltaeh|=|\Deltaee|=0$; possible for all $\Nx$.
The one-particle distribution functions of {\elec}  and {\hole}
are $n\supe(\xi)=\theta(\nu-\xi)$ and $n\suph(\xi)=\theta(-\nu-\xi)$, 
respectively,
where $\theta(x)$ is the step function.

\noindent
II: $|\Deltaeh| \neq 0$, $|\Deltaee|=0$;
possible for $\Nx=0$ and $\nu < |\Deltaeh|$.
$\deltaeh_{\sVk}\neq 0$ for $|\xi| \lesssim |\Deltaeh|$,
and the wave function takes the same form as 
the BCS state,
if
$(e_{\sVk,\sigma}, h_{-\sVk,-\sigma})$ 
is replaced by
$(c_{\sVk,\sigma}, c_{-\sVk,-\sigma})$.
This is the ordinary {\eh} BCS state~\cite{Keldysh-in-BEC}
in nondoped ($\Ne=\Nh$) semiconductors.
The energy cost of adding an electron-like
quasiparticle to this state is 
$E_{\sVk}-\nu$,
where
$E_{\sVk}\equiv \sqrt{\xi_{\sVk}^{2}+|\Deltaeh|^{2}}$.

\noindent
III: $|\Deltaeh|\neq 0$, $|\Deltaee|=0$;
possible for small but finite $\Nx> 0$ and  $\nu > |\Deltaeh|$.
Formally, this solution
(whose wave function is denoted by $|{\rm III} \rangle$)
is 
obtained from solution II 
(whose wave function $|{\rm II} \rangle$)
by adding electron-like quasiparticles
(whose annihilation operator $\epsilon_{\sVk \sigma}$) up to
$E_{\sVk}<\nu$, i.e.,
$| {\rm III} \rangle =
\big( \textstyle\prod'_{\sVk \sigma} \epsilon_{\sVk \sigma}^{\dagger}
\big) |{\rm II} \rangle$,
where
$\textstyle\prod'_{\sVk \sigma}$ denotes the product over the range
$E_{\sVk}<\nu$.
Direct calculation shows that 
$|{\rm III} \rangle =
\big( \textstyle\prod'_{\sVk \sigma} e_{\sVk \sigma}^{\dagger} 
S_{\sVk \sigma}\big)  |{\rm II} \rangle$, 
where
 $S_{\sVk \sigma}$ annihilates an 
($e_{\sVk \sigma},h_{-\sVk,-\sigma}$) pair. 
Therefore, 
$n\supe(\xi)=1$, $n\suph(\xi)=0$, and $\deltaeh_{\sVk}=0$
({\elec} and {\hole} are unpaired) 
in the region $|\xi| < \sqrt{\nu^{2}-|\Deltaeh|^{2}} \equiv \xi'_{\rm F}$.
This unpairing is demonstrated in Fig.~\ref{fig:pair_corr_vs_xi}
by dotted lines, which have discontinuities 
(secondary ``Fermi surfaces'') 
at $\xi=\pm\xi'_{\rm F}$.
We call this solution
the partially unpaired {\eh} BCS state (PU state).
As $\Nx$ increases, $|\Deltaeh|$ diminishes gradually
until it vanishes at a certain value of $\Nx$,
where this solution changes into solution I continuously.

\noindent
IV: $|\Deltaeh|=0$, $|\Deltaee| \neq 0$; possible for all $\Nx$.
Similar to solution I except that
$\deltaee_{\sVk}\neq 0$ for $|\xi-\nu| \lesssim |\Deltaee|$.
This is an ordinary superconductor of electrons.

\noindent
V: $|\Deltaeh|$, $|\Deltaee| \neq 0$
($|\Deltaeh| \gg |\Deltaee|$
because $\lambda^{eh} \gtrsim \lambda^{ee}$); 
possible for small but finite $\Nx> 0$.
Similar to solution III except that
$\deltaee_{\sVk}\neq 0$
if $|\xi \pm \xi'_{\rm F}| \lesssim |\Deltaee|$, i.e.,
the {\ee} pair correlation exists
around the secondary ``Fermi surfaces'' 
(see solid lines in Fig.~\ref{fig:pair_corr_vs_xi}).
This is the only solution where $\Deltaeh$ and $\Deltaee$ 
coexist~\cite{hh_correlation}.
We call this solution
the 
\REV multiply ordered superconducting (MS) \VER
state.

\begin{figure}[tbp]
  \vspace*{0.2cm} \hspace*{0.4cm}
  \epsfxsize=7.5cm
  \epsfbox{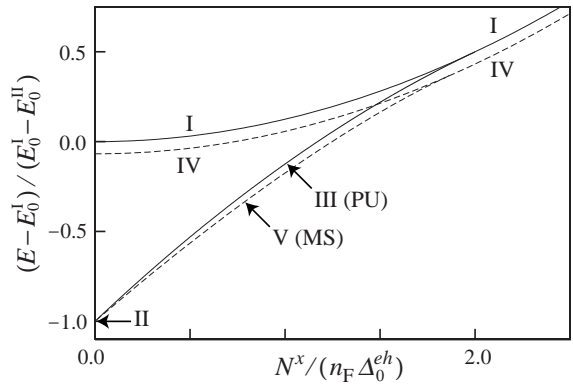}
  \vspace*{0.2cm}
 \caption{
 Plots of $E=\langle H \rangle$
 vs $\Nx\equiv \Ne - \Nh $
 for solutions I--V, 
 when $\lambda\supeh=0.25$ and $\lambda^{ee}=0.20$,
 for a fixed value of 
 $N \equiv (\Ne + \Nh)/2$.
 (The same plot is obtained for any values of $N$ 
 because of the normalization of both axes.)
 $\Deltaeh_{0}$ is the value of $|\Deltaeh|$ at $\Nx=0$
 of solution II.
 $E_{0}^{\rm I}$ and $E_{0}^{\rm II}$ are
 the values of $E$ at $\Nx=0$ for solutions I and II, respectively.
 }
 \label{fig:energy_vs_doping}
\end{figure}
To identify what solution is physically realized, 
we compare their energies 
$E \equiv \langle H \rangle$
for all values of $\Nx \equiv \Ne - \Nh$. 
Note that we compare $\langle H \rangle$ rather than
$\langle {\cal H} \rangle$, 
because the  
natural parameters controlled directly 
by the photoexcitation intensity and the doping
are $\Ne$ and $\Nh$ rather than $\mu^e$ and $\mu^h$
\cite{excitonic_insulator}.
[If we compared $\langle {\cal H} \rangle$, the discussion
would be rather complicated
because 
the relations between $(\Ne, \Nh)$ and $(\mue, \muh)$
are different for different solutions.]
Figure \ref{fig:energy_vs_doping} shows $E$
as a function of $\Nx$, 
where $N \equiv (\Ne + \Nh)/2$.
One recognizes that 
the solution V has the lowest energy for all values of 
$\Nx/(\nF \Deltaeh_{0}) \lesssim 1.86$.
However, care should be taken because its
curve is convex up, which might indicate
a phase separation.
We now show that the solution V is stable 
because {\elec} and {\hole} have opposite  
electrical charges.

To show the stability, 
we first consider the {\em unstable\/} case
where {\elec} and {\hole} denote some 
fermions that have charges of the {\em same\/} sign
(including the neutral case).
All calculations so far are also applicable 
to such a general case (as long as we regard $\Ueh$ and $\Uee$
as given parameters).
%
Since the energy of a single phase of V is larger than 
the average of the energies of  
two phases II and IV, 
the system should undergo a phase separation
into two phases, one with the excess density
$\Nx = 0 < \Nx\subtot$ (phase II) and the other with
$\Nx > \Nx\subtot$ (phase IV), 
where $\Nx\subtot$ denotes the average value of 
$\Nx$ of the total system. As $\Nx\subtot$ is decreased, 
the region(s) of phase IV becomes smaller, until 
the total system turns into a single phase 
II for  $\Nx\subtot=0$.  
In a similar manner, 
one can also show the instability of phase III.
This corresponds to the instability of the Sarma state~\cite{Sarma}
that was discussed in the studies of superconductivity
in a magnetic field. 

On the other hand, the situation is totally different 
if we return to the electron-hole system,
where $e$ and $h$ have opposite charges $-q$  and $q$ ($q>0$),
respectively.
If the phase separation occurred, 
each phase would have global net charge of density
$-q\,\delta N \equiv -q (\Nx-\Nd)$ in phase II, and 
$q\,\delta N$ in phase IV.
The global charge would result in a large cost of 
the long-range part of the Coulomb energy 
$E_{\rm LR}\equiv \langle H_{\rm int}^{\rm LR} \rangle$, 
which we neglected in $H$,
eq.\ (\ref{eq:start_H}).
Taking $E_{\rm LR}$ into account,
the total energy of such a nonuniform state
would be
$E\subtot=\sum_i v_i E_i+E_{\rm LR}+E_{\rm B}$, 
where $v_i$ and $E_i$ denote the volume and the expectation value of $H$ 
for phase $i$, respectively, and $E_{\rm B}$ the boundary energy.
Since $E_{\rm B} \geq 0$, $E\subtot \geq 
\sum_i v_i E_i+E_{\rm LR} \equiv E\subtot'$. 
When the system is in a single phase V, then $E_{\rm LR} =0$ and
$E\subtot$ equals $E$ of phase V of Fig.~\ref{fig:energy_vs_doping}.
On the other hand, when the system is separated into
cells of phases II and IV, one obtains a finite $E_{\rm LR}$, 
\SHMZin
which would be dominated by the electrostatic energy.
We find that 
$E\subtot^{\rm V}$
is smaller than 
$E\subtot^{\prime {\rm II}+{\rm IV}}$ 
($\leq E\subtot^{{\rm II}+{\rm IV}}$),
hence a single phase V should be realized,
except for very small values of $\Nx$
if $\LPippard \gg L_{\rm c}$.
\SHMZout
Here, $\LPippard\equiv v_{\rm F}/(\pi |\Deltaeh|)$ is
the Pippard length of {\eh} pairs,
 which gives the {\em minimum\/} allowable 
size of cells of phase II,
whereas $L_{\rm c} \approx \sqrt{\kappa/(q^{2}\nF)}$
is the {\em maximum\/} allowable size 
of cells of phase II,
below which the energy decrease by the phase
separation overcomes the increase of $E_{\rm LR}$, 
where $\kappa$ denotes the dielectric constant of the 
semiconductor.
The condition $\LPippard \gg L_{\rm c} $ is {\em always\/} satisfied 
at high densities, 
i.e., when 
$r_{\rm s} \equiv [3/(4\pi N)]^{1/3} m q^{2}  / (4\pi \kappa \hbar^{2})
\ll 1 $ and 
$N\supe \approx N\suph$ ($\approx N$).
In fact, in this case  
we may assume a screened Coulomb interaction
for  $\Veh$ 
and can easily show that
$ \LPippard / L_{\rm  c} > 10^{8}$ 
for {\em any\/} $r_{\rm s}\lesssim 1$. 
Therefore, at high densities ($r_{\rm s} \ll 1$) the 
phase V is always stable against phase separation.
Moreover, considering the large value $10^8$, we may 
extend this conclusion to densities 
where $r_{\rm s} \approx 1$~\cite{rs>1_case}.
The condition $r_{\rm s} \lesssim 1$ can be realized
when, e.g., $m=0.1 m\supe_{0}$, $\kappa=10 \kappa_{0}$
and $N=10^{19}\,{\rm cm}^{-3}$,
for which we obtain $r_{\rm s}=0.54$,
where  $m\supe_{0}$ is the free electron mass and
$\kappa_{0}$ the dielectric constant of vacuum.

\begin{figure}[t]
  \vspace*{0.0cm}\hspace*{-0.0cm}
  \epsfxsize=8.0cm
  \epsfbox{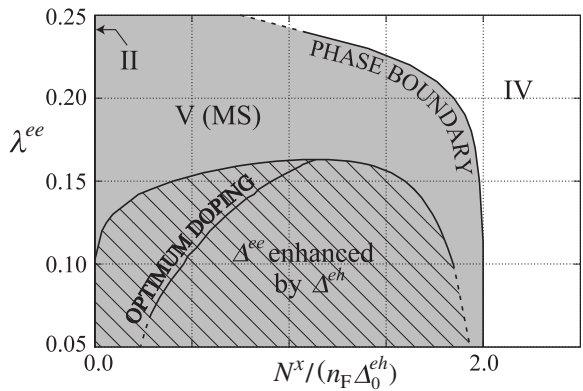}
  \vspace*{0.2cm}
 \caption{
 Phase diagram of a high-density {\eh} plasma 
 for $\lambda\supeh=0.25$.
 The 
 \REV MS \VER
 phase is shadowed.
 $|\Deltaee|$ is enhanced in the hatched region,
 where $|\Deltaee|$ takes maximum values 
 with respect to $\Nx$ on the line denoted
 as optimum doping.
 The vertical line at $\Nx=0$ corresponds to phase II.
 Dashed lines are guides for the eye (not calculated).
 }
 \label{fig:phase_diagram}
\end{figure}
\begin{figure}[t]
 \vspace*{0.0cm}
 \hspace*{0.0cm}
 \epsfxsize=7.5cm
 \epsfbox{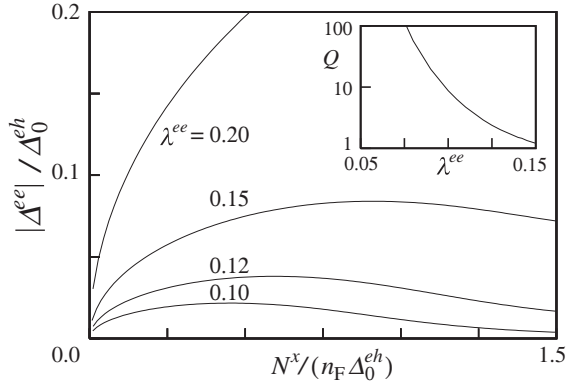}
 \vspace*{0.2cm}
 \caption{
 The $\Nx$ dependence of $|\Deltaee|$ in 
 the 
 \REV MS \VER 
 phase for $\lambda\supeh=0.25$. 
 The inset shows the enhancement factor $Q$ vs $\lambda\supee$.
 }
 \label{fig:order_param_vs_doping}
\end{figure}

By performing calculations of the energies of various solutions
as a function of $\lambda^{ee}$ and $\Nx$, 
we obtain the 
phase diagram shown in Fig.~\ref{fig:phase_diagram}.
Phase V 
(\REV MS \VER phase), 
for which $|\Deltaeh| \gg |\Deltaee| > 0$,
is realized in the shadowed region.
This phase changes continuously into 
phase II as $\Nx \to 0$.
On the other hand,
this phase changes into phase IV as $\Nx$ increases,
because then the Fermi energies of {\elec} and {\hole}
become separate, which prevents {\eh} pairing.
To study the $\Nx$ dependence in more detail, 
we plot $|\Deltaee|$ in 
the \REV MS \VER phase 
in Fig.~\ref{fig:order_param_vs_doping}.
One sees that $|\Deltaee|$ strongly depends on $\Nx$
and takes a maximum value $\Deltaee\subopt$ at an optimum 
$\Nx$ ($\equiv\Nxopt$),
although $\Ne = N + \Nx/2$ is almost constant 
(because $\Nx \ll N$).
This is in marked contrast with the single-component case, 
where the BCS theory gives 
$|\Deltaee|=2\omega_{\rm c} \exp(-1/\lambda^{ee})
\equiv \Deltaee\subBCS$,
which is constant when 
$\Ne$ and $\lambda\supee$ are constant.
To compare $\Deltaee\subopt$ with $\Deltaee\subBCS$,
we plot the enhancement factor 
$Q \equiv \Deltaee\subopt/\Deltaee\subBCS$
in the inset of Fig.~\ref{fig:order_param_vs_doping}.
Since $\Deltaee\subBCS$ is the magnitude of $|\Deltaee|$
which would arise if there were no {\eh} interaction,
the fact $Q>1$ implies that
{\em $|\Deltaee|$ is enhanced by the presence of $|\Deltaeh|$\/}.
It is found that {\em an enhancement by one order of magnitude or 
larger\/} occurs
as $\lambda^{ee}$ becomes smaller. 
As $\lambda^{ee} \to 0$, in particular, 
$|\Deltaee|$ diminishes much more slowly than 
$\Deltaee\subBCS$, resulting in a very large enhancement factor $Q$.
Since the superconducting transition temperature $T_{\rm c} \supee \propto$ 
$|\Deltaee|$, the 
\REV MS \VER phase
should have a $T_{\rm c}\supee$ which is much higher than 
the BCS transition 
temperature~\cite{T_c_in_photo-excited_semiconductors}.
The parameter region in which this enhancement occurs 
is hatched in Fig.~\ref{fig:phase_diagram}, where
the optimum doping line is also shown.


\begin{figure}[t]
 \vspace*{0.0cm}
 \hspace*{0.5cm}
 \epsfxsize=7cm
 \epsfbox{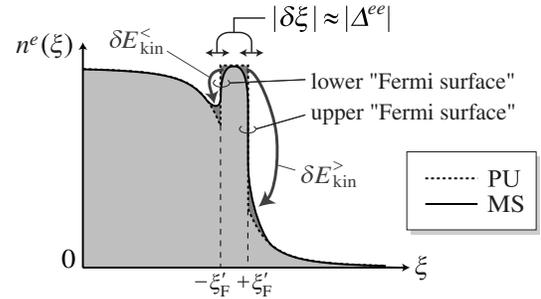}
 \vspace*{0.2cm}
 \caption{
 Schematic illustration of the enhancement mechanism.
 The electron distributions $n^e(\xi)$ 
 for the PU state (dotted line) 
 and the 
 \REV MS \VER  state (solid line)
 are shown.
 }
 \label{fig:mechanism_of_enhancement}
\end{figure}

%
%
%
%
%
To reveal the physical mechanism leading to 
the enhancement of $|\Deltaee|$, we reexamine 
the $e$ distributions of the 
\REV MS \VER 
and PU states
of Fig.~\ref{fig:pair_corr_vs_xi}(a),
which is schematically plotted in 
Fig.\ \ref{fig:mechanism_of_enhancement}.
In the PU state,
for which $|\Deltaee|=0$,
excess electrons are concentrated in the region
$|\xi|\le \xi'_{\rm F}$, and 
the secondary (lower and upper) ``Fermi surfaces''
appear at $\xi = \pm \xi'_{\rm F}$.
In the MS state,
electrons form {\ee} pairs to 
benefit from the attractive {\ee} interaction
and consequently
$\deltaee_{\sVk}$ develops around those ``Fermi surfaces''
[Fig.~\ref{fig:pair_corr_vs_xi}(d)].
However, this pairing 
necessarily broadens
the ``Fermi surfaces'',
resulting in increase of the kinetic energy $E_{\rm kin}$.
Hence, the total energy is minimized for 
an optimum $\deltaee_{\sVk}$.
Although this mechanism is similar to 
the BCS instability in ordinary superconductors, 
there is an essential difference:
In the BCS state,
the broadening of the Fermi surface of width $\delta \xi$
increases the kinetic energy by
$\delta E_{\rm kin} \approx \nF |\delta \xi|^{2} 
\approx \nF |\Deltaee|^{2}$.
In the
\REV MS \VER 
state, on the other hand, 
the broadenings of the
lower and upper ``Fermi surfaces'' change
the kinetic energy by
$\delta E_{\rm kin}^{<} \approx - \nF |\Deltaee|^2$
and
$\delta E_{\rm kin}^{>} \approx + \nF |\Deltaee|^2$,
respectively
[when $\lambda\supee$ (thus $|\delta \xi|$) is small].
These two contributions cancel to give a relatively small increase in the
kinetic energy
$\delta E_{\rm kin} 
= \delta E_{\rm kin}^{<} + \delta E_{\rm kin}^{\rm >}
$.
This is the origin of the huge enhancement factor $Q$ for
small $\lambda\supee$.
Since the cancellation is not perfect, 
$|\Deltaee|$ of course takes a finite value.
As $\lambda\supee$ is increased, the cancellation becomes 
less complete, and $Q$ is reduced.
%
%
%
%
%
The existence of an optimum doping can be understood in 
a similar manner:
If $\Nx$ is too small, no room for broadening is left
between the two ``Fermi surfaces''.
Conversely, if $\Nx$ is too large,
the magnitude of the jump of $n\supe(\xi)$
at the two ``Fermi surfaces'' differs,
which makes the cancellation of contributions from
the two ``Fermi surfaces'' less perfect,
resulting in a smaller $|\Deltaee|$.
Therefore, $|\Deltaee|$ takes a maximum value at
an intermediate value of $\Nx$.


Finally, we point out that 
although we have assumed the {\eh} BCS state 
in semiconductors, 
the same mechanism 
for the enhancement of the {\ee} correlation upon doping 
may be expected also in other ordered phases 
of other materials,
if the electron distribution without the {\ee} correlation is 
similar to the dotted line of Fig.\ \ref{fig:mechanism_of_enhancement}.








\begin{thebibliography}{99}

\bibitem[*]{address_Shigemori}
	   E-mail: shige@asone.c.u-tokyo.ac.jp
\bibitem[**]{address_Brandes}
	   Present address: I. Institut f\"ur Theoretische Physik,
	   Jungiusstra{\ss}e 9, D-20355 Hamburg, Germany.

 \bibitem{Keldysh-in-BEC}
	 L. H. Keldysh: {\it Bose--Einstein Condensation\/},
	 ed. A. Griffin {\it et al.\/}
	 (Cambridge, Cambridge, 1995) p.246.

 \bibitem{EHL}
	 R. S. Markiewicz, J. P. Wolfe and C. D. Jeffries:
	 Phys. Rev. B {\bf 15} (1977) 1988.
	 
 \bibitem{condensation-1}
	 E. Fortin, S. Fafard and A. Mysyrowicz:
	 Phys. Rev. Lett. {\bf 70} (1993) 3861.

 \bibitem{condensation-2}
	 L.V. Butov {\it et al.\/}:
	 Phys. Rev. Lett. {\bf 73} (1994) 304.

	 
 \bibitem{excitonic_insulator}
\SHMZin
We here use the word 
``{\eh} BCS state'' to avoid confusion with 
the ``excitonic insulator'' in semimetals, 
which has the same order parameter.
The former is realized in semiconductors, for which 
$E_{\rm g}\gg|\Ueh|$, so that 
the rigid-band approximation (RBA) is good and 
$N^e$ and $N^h$ are independent of $\Ueh$.
For the latter, on the other hand, 
$|E_{\rm g}| \sim |\Ueh|$,
so that 
the RBA breaks down and
$N^e$ and $N^h$ strongly depend on $\Ueh$.

	 
 \bibitem{exciton_mechanism-1}
	 W. A. Little: Phys. Rev. {\bf 134} (1964) 1416.

 \bibitem{exciton_mechanism-2}
	 D. Allender, J. Bray and J. Bardeen:
	 Phys. Rev. B
	 {\bf 7} (1973) 1020.


 \bibitem{previousEI-1}
	 V. S. Babichenko and M. N. Kiselev:
	 Physica {\bf C209} (1993) 133.
	 
 \bibitem{previousEI-2}
	 A. A. Gorbatsevich and
	 Yu. V. Kopaev:
	 Pis'ma Zh. Eksp. Teor. Fiz. {\bf 51} (1990) 327 
	 [JETP Lett. {\bf 51} (1990) 375].
	 

 \bibitem{McMillan}
	 W. L. McMillan: Phys. Rev. {\bf 167} (1968) 331.
	 
\bibitem{migdal}
For example, the Migdal approximation is {\em only\/} justified for the electron-phonon interactions in metals.



	 
	 %

 \bibitem{different_cutoffs}
	 We take the same cutoff for both $\Ueh$ and $\Uee$,
	 because we found that the final results are insensitive to the
	 cutoff of $\Uee$.

 \bibitem{two-band-sc}
	 G. Gladstone, M. A. Jensen and J. R. Schrieffer:
	 {\it Superconductivity\/},
	 ed. R. D. Parks
	 (Marcel Dekker, New York, 1969).
	 
 \bibitem{Moulopoulos-Ashcroft}
	 K.\ Moulopoulos and N.\ W.\ Ashcroft:
	 Phys.\ Rev.\ Lett.\ {\bf 66} (1991) 2915 
	 treated nondoped systems with similar equations.


	 
	 
	 
	 
	 
	 
 \bibitem{hh_correlation}
	 When an {\hh} interaction $\Uhh<0$ is included 
	 in ${\cal H}$ of eq.\ (\ref{eq:start_H}),
	 {\hh} order parameter $|\Deltahh|$ also becomes finite.
	 
 \bibitem{Sarma}
	 G. Sarma:
	 J. Phys. Chem. Solids {\bf 24} (1963) 1029.

	 

	 
	 
 
 \bibitem{rs>1_case}
	 When $r_{\rm s}>1$, 
	 our assumption $\omega_{\rm c}\ll\mu$ no longer holds,
	 and therefore the stability of 
	 the \REV MS \VER state
	 for $r_{\rm s}>1$ is
	 beyond the scope of this Letter.
	 
	 


 \bibitem{T_c_in_photo-excited_semiconductors}
	 If $\lambda^{ee}$ is very small,
	 $T_{\rm c}^{ee}$ is low even when $Q\gg 1$.


\end{thebibliography}
\end{document}